\documentclass[trackchanges,modern]{aastex701}

\usepackage{float}
\usepackage{xspace}
\usepackage{enumitem}

\graphicspath{{paper_figs_small/}}

\newcommand{\bea}{\begin{eqnarray} }
\newcommand{\eea}{\end{eqnarray}}

\begin{document}

\title{Vaulting the barrier: An intrinsic mechanism to fuel the gas beyond the nuclear ring into the central region of barred galaxies}

\author[]{Kotaro Kobayashi}
\affiliation{Kagoshima University, Graduate School of Science and Engineering, Kagoshima 890-0065, Japan}
\email{k.kobayashi@example.jp}

\author[0009-0008-8432-7460]{Naomichi Yutani}
\affiliation{Kobe University, Kobe, 657-8501, Japan}
\email{yutaninm@gmail.com}

\author[0000-0001-8226-4592]{Takayuki R. Saitoh}
\affiliation{Kobe University, Kobe, 657-8501, Japan}
\email{saitoh@people.kobe-u.ac.jp}

\author[0000-0002-2154-8740]{Junichi Baba}
\affiliation{Kagoshima University, Graduate School of Science and Engineering, Kagoshima 890-0065, Japan}
\affiliation{National Astronomical Observatory of Japan, Japan}
\email{baba@example.jp}

\author[0000-0002-8779-8486]{Keiichi Wada}
\affiliation{Kagoshima University, Graduate School of Science and Engineering, Kagoshima 890-0065, Japan}
\affiliation{Ehime University, Research Center for Space and Cosmic Evolution, Matsuyama 790-8577, Japan}
\affiliation{Hokkaido University, Faculty of Science, Sapporo 060-0810, Japan}
\email[show]{wada@astrophysics.jp}
\correspondingauthor{Keiichi Wada}
\begin{abstract}
{
Gas delivery to galactic centers powers nuclear starbursts and active galactic nuclei (AGNs), yet bar-driven inflow is generally expected to stall in a nuclear ring a few hundred parsecs across. Using three-dimensional Lagrangian hydrodynamic simulations in a fixed barred potential, we identify a bypass channel in which a fraction of the inflowing gas acquires vertical momentum, vaults across the ring, and reaches the inner few tens of parsecs. This pathway is absent in two-dimensional calculations, which instead predict long-lived stagnation at the ring. We find that the circumnuclear material within $\sim 50$~pc originates from gas initially located outside the ring ($\gtrsim 300$~pc), rather than from secondary inflow out of the ring itself. Successful delivery requires both a sufficiently large vertical excursion, $|z| \sim 100$~pc before encountering the ring, and substantial loss of azimuthal angular momentum $L_z$. 
The resulting inflow is organized rather than chaotic: center-reaching trajectories are confined to a limited spatial region set by the scale height of the ring gas. Most bar-driven gas still accumulates near the resonance and fuels star formation in the nuclear ring, but the vaulting stream selects a modest yet sufficient fraction that penetrates to the circumnuclear disk. These results suggest that intrinsically three-dimensional gas motions help link nuclear starbursts, AGN fueling, and the frequent misalignment of nuclear disks with respect to their host galaxies.}

\end{abstract}

\keywords{\uat{Galactic dynamics}{591}, \uat{Galactic bars}{568}, \uat{Hydrodynamical simulations}{767}, \uat{Gas dynamics}{622}, \uat{Accretion}{14}}

\section{Introduction}

Gas inflow to galactic centers is essential for bulge formation, nuclear starbursts, and the growth of supermassive black holes (SMBHs). The observed correlation between SMBH mass and host-galaxy spheroid mass suggests a close connection between galactic-scale gas supply and nuclear activity \citep{Rees1984, 2010MNRAS.407.1529H, KormendyHo2013, combes_book2021}. Transporting gas from kiloparsec scales to the parsec-scale vicinity of an SMBH, however, requires the removal of a vast amount of angular momentum. On galactic scales, stellar bars drive gas inflow \citep{Athanassoula1992}, producing spiral structures near the inner Lindblad resonance (ILR) on sub-kpc scales \citep{WadaHabe1992, Wada1994}. Offset shocks associated with dust lanes further promote transport down to several hundred parsecs.

However, bar-driven fueling often stalls when gas accumulates in ring-like structures \citep{ButaCombes1996, Regan2003, Regan2004, Kim-Teuben2012, Emsellem2015}. Sustained inflow beyond such rings therefore appears to require additional mechanisms, such as self-gravitational collapse \citep{WadaHabe1992}, stellar feedback \citep{DiMatteo2005}, or shocks in nuclear spirals \citep{EnglmaierShlosman2000, Maciejewski2004}. The ``bars-within-bars'' hypothesis \citep{Shlosman1989, Shlosman1990, maciejewski2002} proposes a hierarchical cascade in which secondary instabilities drive gas inward to $\sim 10$ pc, although this process remains difficult to demonstrate in realistic settings \citep[e.g.,][]{Wada1995}. Higher-resolution simulations and observations have further suggested that non-axisymmetric $m=1$ modes or supernova-driven turbulence may also contribute to inward transport \citep{Wada+09, HopkinsQuataert2011}.

High-resolution Atacama Large Millimeter/submillimeter Array (ALMA) observations of nearby galaxies, particularly in surveys such as NUGA and GATOS, suggest that molecular tori in the central tens of parsecs are statistically decoupled in orientation from the host galaxy bar and disk \citep{Combes2019, GarciaBurillo2021}. These results imply that the fueling process responsible for assembling such dense nuclear gas was not simply a continuous, two-dimensional inflow within the galactic disk. Simulations by \citet{Emsellem2015} showed that stellar feedback can eject gas vertically, which then falls back with reduced angular momentum and may form polar disks. Recent studies have also highlighted ``overshoot'' orbits, in which gas flowing along the bar passes through the central region and then returns outward \citep{Sormani2020}. \citet{Chaves-Velasquez2024-ty} found that gas supplied to the Central Molecular Zone (CMZ) of our Galaxy can first expand vertically owing to shock-induced heating and subsequently settle into a thinner layer.

In this paper, we revisit the interruption of mass supply by nuclear rings in a bar potential using high-resolution three-dimensional smoothed particle hydrodynamics (SPH) simulations. Through a detailed Lagrangian analysis, we identify a new fueling channel in which gas ``vaults'' the nuclear ring through modest vertical excursions. 

This paper is organized as follows. In Section 2, we describe the galactic model, numerical methods, and the definitions of the ``nuclear ring'' and ``circumnuclear disk'' adopted in this paper. In Section 3, we present the simulation results. To clarify the physical origin of the new fueling mode, we perform a pseudo-two-dimensional calculation and compare it with the three-dimensional results
under the isothermal approximation (Section 3.2), and we analyze the underlying mechanism in Section 3.3. In Section 4, we discuss the relation of our results to recent numerical studies and compare them with observations. Section 5 summarizes our conclusions.

\section{Model and Methods}

In this study, the stellar and dark-matter components are represented by externally imposed potentials, and the gas motion is solved with the SPH method.
The simulations solve gravity and hydrodynamics and also include heating, cooling, and star formation.

\subsection{Galaxy Model}

We prescribe the stellar disk, dark-matter halo, bulge, supermassive black hole (SMBH),
and the bar as external potentials.
We adopt the Miyamoto--Nagai disk \citep{miyamoto1975} for the stellar disk:
\begin{equation}
\rho_\mathrm{disk}(R,z)
=
\frac{b^2 M_\mathrm{disk}}{4\pi}
\frac{
a R^2 + (a + 3 \sqrt{z^2+b^2})(a+\sqrt{z^2+b^2})^2
}{
\left[R^2 + (a+\sqrt{z^2+b^2})^2\right]^{5/2}
(z^2+b^2)^{3/2}
},
\label{eq: MN-disk}
\end{equation}
where $R$ and $z$ denote the cylindrical radius and vertical coordinate, $M_\mathrm{disk}$ is the disk mass, and $a$ and $b$ are the radial and vertical scale lengths.
We set
$M_\mathrm{disk} = 5.5 \times 10^{10}\,M_\odot$,
$a = 1.5~\mathrm{kpc}$,
and
$b = 400~\mathrm{pc}$.
For the dark-matter halo,  we assume an NFW profile \citep{navarro1996}:
\begin{equation}
\rho_\mathrm{halo}(r)
=
\frac{\rho_0}{(r/r_s)(1+r/r_s)^2},
\end{equation}
where $r$ is the spherical radius, $r_s$ is the scale radius, and $\rho_0$ is the characteristic halo density.
We adopt
$\rho_0 = 6.68 \times 10^{6}\, M_\odot\, \mathrm{kpc}^{-3}$
and
$r_s = 20~\mathrm{kpc}$.
The bulge follows the Hernquist profile \citep{hernquist1990}:
\begin{equation}
\rho_\mathrm{bulge}(r)
=
\frac{M_\mathrm{bulge}}{2 \pi}
\frac{a}{r(r+a)^3}, 
\end{equation}
where $M_\mathrm{bulge}$ and $a$ are the bulge mass and scale radius.
We set
$M_\mathrm{bulge} = 1 \times 10^{9}~M_\odot$
and
$a = 1~\mathrm{kpc}$.
For the SMBH,  the Plummer potential is assumed \citep{plummer1911}:
\begin{equation}
\rho_\mathrm{BH}(r)
=
\frac{3 M_\mathrm{BH}}{4 \pi \epsilon_\mathrm{BH}^3}
\left(
1 + \frac{r^2}{\epsilon_\mathrm{BH}^2}
\right)^{-5/2}.
\end{equation}
Here $M_\mathrm{BH}$ is the SMBH mass and $\epsilon_\mathrm{BH}$ is the gravitational softening length.
We adopt
$M_\mathrm{BH} = 5 \times 10^{8}~M_\odot$
and
$\epsilon_\mathrm{BH} = 1~\mathrm{pc}$.

The axisymmetric stellar component of the galactic potential is
\begin{equation}
\Phi_0 = \Phi_{\mathrm{disk}} + \Phi_{\mathrm{bulge}},
\label{eq: potential}
\end{equation}
where $\Phi_{\mathrm{disk}}$ and $\Phi_{\mathrm{bulge}}$ denote the gravitational potentials of the stellar disk and bulge, respectively.
We adopt the Sanders bar potential \citep{sanders1976} for the bar:
\begin{equation}
\Phi_{\mathrm{bar}}(R,\phi)
=
\epsilon\, f(R)\, \Phi_0 \cos(2\phi),
\label{eq: bar_potential}
\end{equation}
where
\begin{equation}
f(R)
=
\frac{a R^2}{\left(a^2 + R^2\right)^{3/2}},
\end{equation}
Here $\epsilon$ and $a$ are the bar strength and core radius
\footnote{{ 
For the potentials given by Eqs. \ref{eq: potential} and \ref{eq: bar_potential}, the underlying density is always positive within the region $|z| < 1$ kpc for the adopted parameters.
 Furthermore, the gas motion is confined to the region $|z| < 300$ pc.}}
We choose
$a = 2~\mathrm{kpc}$
and
$\epsilon = 0.12$.
The bar pattern speed is set to
$\Omega_p = 24~\mathrm{km\,s^{-1}\,kpc^{-1}}$.
The resulting rotation curve and angular-velocity profile are comparable to those in \citet{fukuda2000},
and the inner Lindblad resonances are located at $R = 0.72~\mathrm{kpc}$, $1.56~\mathrm{kpc}$, and $2.9~\mathrm{kpc}$, while the corotation radius is at $R = 8.85~\mathrm{kpc}$.

Following \citet{dehnen2000}, we gradually increase the bar strength $\epsilon$ to avoid numerical instabilities immediately after the start of the calculation:
\begin{equation}
\epsilon(t) =
\left\{
\begin{array}{ll}
0, & (t < 50~\mathrm{Myr}), \\
\epsilon \left(
\frac{3}{16}\eta^5
- \frac{5}{8}\eta^3
+ \frac{15}{16}\eta
+ \frac{1}{2}
\right),
& (50~\mathrm{Myr} \le t < 150~\mathrm{Myr}), \\
\epsilon, & (t \ge 150~\mathrm{Myr})
\end{array}
\right.
\end{equation}
where
\begin{equation}
\eta = 2\,\frac{t - 50~\mathrm{Myr}}{100~\mathrm{Myr}} - 1
\end{equation}
\citep{dehnen2000}.
We fix $\epsilon=0$ for the first $50~\mathrm{Myr}$, ramp it to $0.12$ between $50$ and $150~\mathrm{Myr}$, and analyze only the time interval between $150$ and $200~\mathrm{Myr}$ when the bar is fully developed.

The initial gas-disk density distribution is
\begin{equation}
\rho(R,z)
=
\frac{\Sigma_0}{\sqrt{2\pi}\sigma_z}
\exp\left(
-\frac{R}{R_d}
-\frac{z^2}{2\sigma_z^2}
\right),
\end{equation}
where $\Sigma_0$ is the central surface density,
$R_\mathrm{d}$ is the scale length of the gas disk,
and $\sigma_z$ is the vertical dispersion.
We adopt
$\Sigma_0 = 50~M_\odot\,\mathrm{pc}^{-2}$,
$R_\mathrm{d} = 3~\mathrm{kpc}$,
and
$\sigma_z = 10~\mathrm{pc}$.
The disk is truncated at a radius of $3.5~\mathrm{kpc}$ and its initial temperature is $10^{4}~\mathrm{K}$.

\subsection{Numerical Method}

We investigate the gas dynamics with the Lagrangian smoothed particle hydrodynamics (SPH) scheme.
Specifically, we employ the parallel Tree+GRAPE $N$-body/SPH code ASURA \citep{saitoh2008,saitoh2013}.
In addition to hydrodynamics and self-gravity, the code includes heating, radiative cooling, and star formation.
We assume a uniform far-UV radiation field of $G_0 = 1.7$, comparable to that in the Solar neighborhood, and we neglect supernova feedback for simplicity.
Star formation is modeled by stochastically converting SPH particles that satisfy the criteria
$T < 100~\mathrm{K}$,
$n_{H} > 100~\mathrm{cm^{-3}}$,
and
$\nabla \cdot \mathbf{v} < 0$
into star particles, which subsequently behave as gravitating sources.
For comparison we also run isothermal models without star formation (see Section \ref{sec:2d-isothermal}).

{ Note on star formation. The primary aim of this work is to investigate the dynamical behavior of gas, particularly the vaulting mechanism. The star formation recipe adopted here is therefore not intended to model realistic star formation, but rather to prevent the excessive formation of dense clumps via gravitational collapse, thereby maintaining numerical stability. For this reason, we adopt a relatively large gravitational softening length (30 pc) and do not include stellar feedback. The effects of stellar feedback will be explored in a future study.}

We impose an inner boundary at $5~\mathrm{pc}$, and SPH particles that enter this region are accreted onto the SMBH, which is fixed at the origin.
Accretion takes place when (1) the specific energy of a particle with respect to the SMBH is negative and (2) its specific angular momentum falls below the Keplerian value.

The simulations start with $5 \times 10^{6}$ SPH particles and use a gravitational softening length of $30~\mathrm{pc}$.
The corresponding mass resolution is $184~M_\odot$.

\subsection{Definition of CND and the nuclear ring}
\label{sec:2.4}
To understand the underlying physics of the results (see Section \ref{sec:vertical}), 
we need to define the region
$r=\sqrt{x^{2}+y^{2}+z^{2}}<50~\mathrm{pc}$
as the CND and quantify the gas inflow into this volume.
Because the ring radius changes with time, we determine it at each snapshot from the gas surface-density distribution.
We define $R_x$ and $R_y$ as the radii of the surface-density maxima along the bar major and minor axes, respectively, and compute the nuclear-ring radius as
$
  R_{\rm ring} = \sqrt{R_x R_y}.
$
When evaluating the radius of maximum surface density we mask the CND region ($r<50~\mathrm{pc}$).

The nuclear-ring region is defined by $0.7R_{\rm ring} < R < 1.3R_{\rm ring}$, where $R = \sqrt{x^{2} + y^{2}}$.
$R_{\rm ring}$ is defined as the time-averaged value over $t = 150$--$200~\mathrm{Myr}$ (with 1 Myr sampling).
Within this region we use the mass-weighted mean specific angular momentum and the \textit{half-mass height} to characterize the ring angular momentum and scale height.

\subsection{Decomposition of Gravitational and Hydrodynamic Torques}
\label{sec:2.5}

We decompose the torque that changes the angular momentum of each SPH particle at every timestep into gravitational and hydrodynamic contributions.
The hydrodynamic torque is further separated into the components generated by the pressure gradient and by the artificial viscosity, and these contributions are analyzed in detail in Section~\ref{sec:shock}.
The contribution of each torque component to the angular momentum $L_i$ of particle $i$ is evaluated concurrently with the time integration via
\begin{equation}
L_i(t^{n+1}) - L_i(t^{n})
= \int_{t^{n}}^{t^{n+1}} \tau_i(t)\,\mathrm{d}t
\end{equation}

\begin{equation}
\simeq \frac{\Delta t}{2}\,\tau_i(t^{n})
+ \frac{\Delta t}{2}\,\tau_i(t^{n+1})
\end{equation}
where $L_i(t)$ and $\tau_i(t)$ are the $i$th components of the angular-momentum and torque vectors at time $t$.

\section{Results}

\subsection{Gas Inflow into the CND: Vaulting over the Nuclear Ring}

\begin{figure}[htb!]
\centering
\includegraphics[width=100mm]{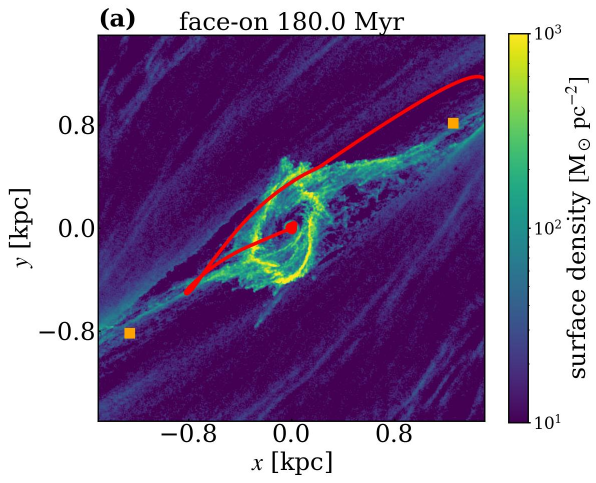}
\par\bigskip
\includegraphics[width=100mm]{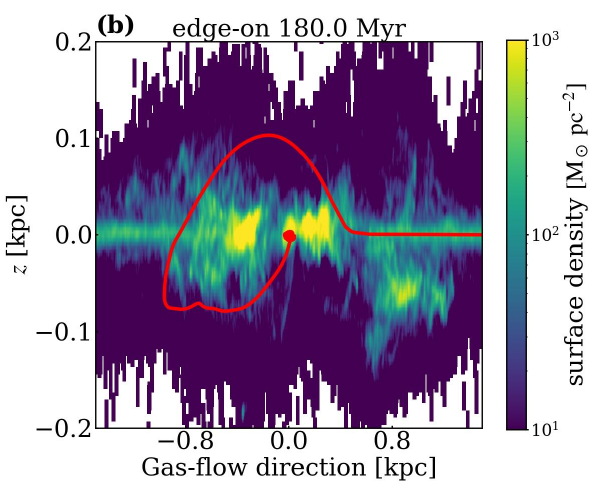}
\caption{
(a) Face-on gas surface-density map at $t = 180~\mathrm{Myr}$ with the orbit of the representative particle that reaches the CND (red curve). The $x$-axis aligns with the bar major axis and the bar rotates counterclockwise.
(b) Slice along the line connecting the endpoints of the orange square in panel (a). The horizontal and vertical axes correspond to the direction along the gas flow and perpendicular to the disk, respectively. The $z$-direction is vertically exaggerated.  {  See also the attached movie.}
}
\label{fig:figure1}

\end{figure}

Figure~\ref{fig:figure1} overlays the trajectory of a representative gas particle that eventually reaches the CND region ($r < 50~\mathrm{pc}$) on the gas surface-density map at $t=180~\mathrm{Myr}$. The red solid line traces the orbit taken between 150 and 180 Myr. The face-on map highlights the nuclear ring and the elongated, bar-aligned overdensities that correspond to the offset dust lanes commonly seen in nearby barred galaxies. The selected particle is lifted out of the mid-plane near the ring, then encounters the opposite-side shock in the dust-lane region, allowing it to get over the ring and fall into the galactic center. 
We refer to this fueling process as the ``vaulting mechanism.'' Vertical motions out of the galactic plane are induced by oblique shocks in the dust lanes and the nuclear ring (see Section~\ref{sec:vertical}  in detail).


\subsection{Comparison between Three-dimensional and Pseudo-2D Simulations}
\label{sec:2d-isothermal}
\begin{figure}[htb!]
\centering
\includegraphics[width=\linewidth]{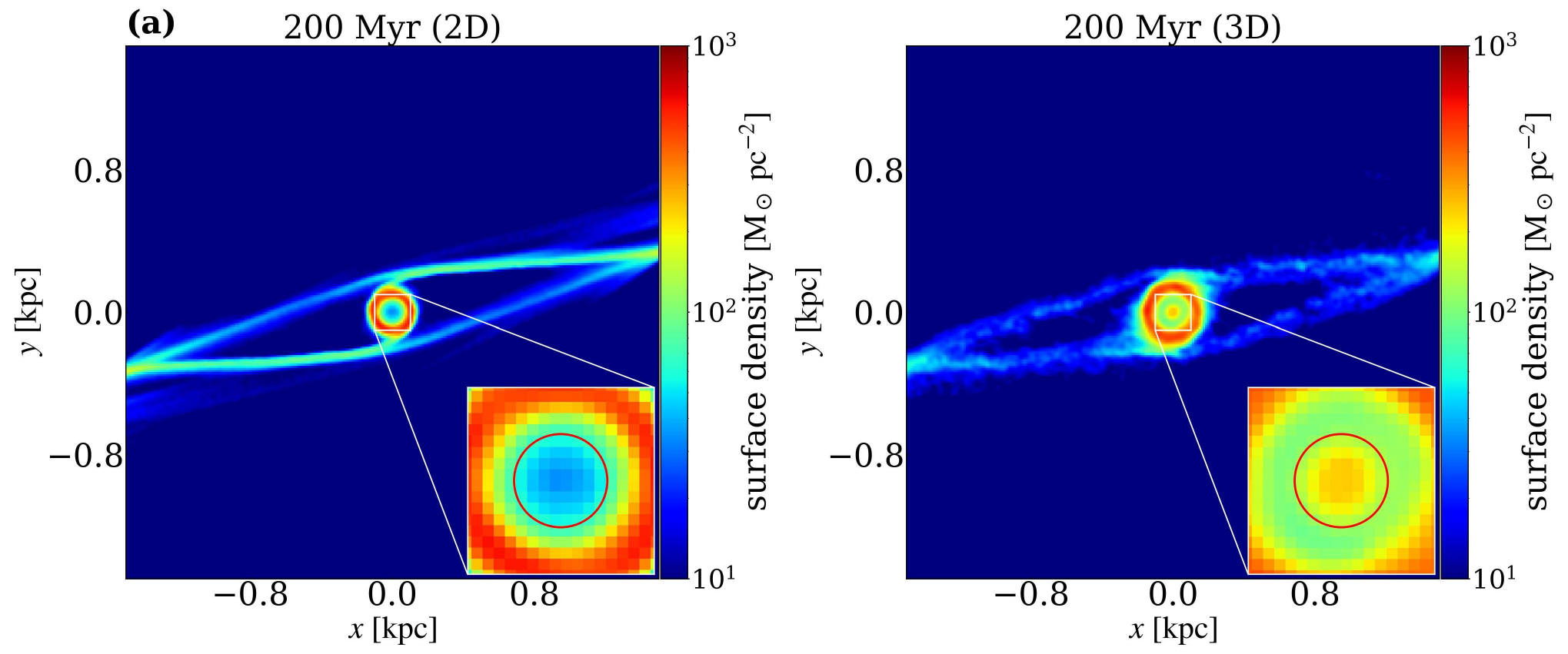}
\par\bigskip
\includegraphics[width=\linewidth]{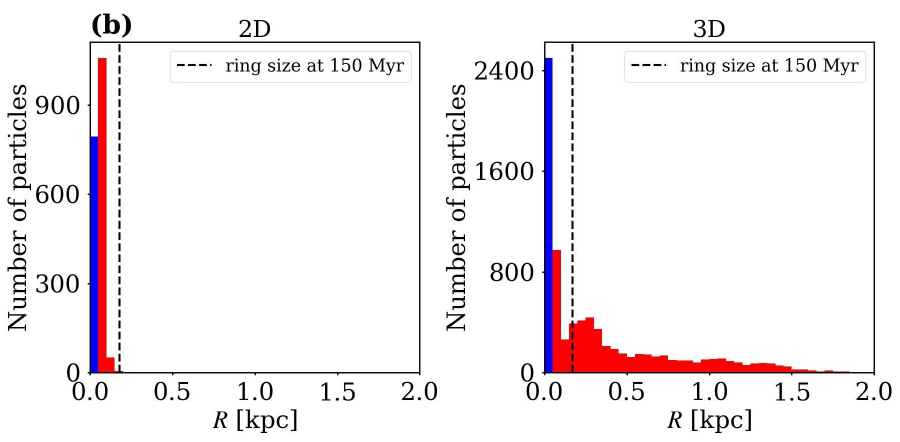}

\caption{
(a) Surface-density maps at $t=200~\mathrm{Myr}$ for the pseudo-2D (left) and 3D, isothermal (right) models. The red circle marks the CND region ($r<50~\mathrm{pc}$).
(b) Radial distributions at $t=150~\mathrm{Myr}$ of gas particles that are in the CND at $t=200~\mathrm{Myr}$, obtained by tracing those particles back to $t=150~\mathrm{Myr}$. Blue shows particles already inside the CND at $t=150~\mathrm{Myr}$, while red shows particles that were outside the CND at $t=150~\mathrm{Myr}$ and entered it during $150$--$200~\mathrm{Myr}$. The dashed line indicates the ring radius at $t=150~\mathrm{Myr}$. In the pseudo-2D model, the CND is built almost entirely from gas already located inside the ring, whereas in the 3D, isothermal model a substantial fraction of the CND gas originates from outside the ring.
}

\label{fig:figure2}

\end{figure}

The ``vaulting mechanism'' described above is intrinsically three-dimensional, and we demonstrate the importance of the vertical motion even for geometrically thin disks by comparing with an experimental ``2D'' simulation, where
the vertical motion is artificially restricted using the same SPH code. To isolate the key differences, this ``pseudo-2D" calculation adopts an isothermal assumption and omits the detailed heating/cooling and star-formation physics. The main setup matches the fiducial run shown in the previous section except for the following changes:
\begin{itemize}
\item The initial gas distribution is confined to a disk plane, and the vertical acceleration is forced to zero so that the motion remains two-dimensional.
\item The gas is isothermal at $10^{4}~\mathrm{K}$ and no star particles form.
\item The particle number is $10^{6}$, which yields an initial central surface density of $\Sigma_0 = 10~M_\odot\,\mathrm{pc}^{-2}$, i.e., 20\% of the fiducial model.
\end{itemize}
{ We compare this pseudo-2D model with a 3D model that also adopts the isothermal condition (hereafter the ``isothermal-3D'' model). In both models, gas self-gravity is included.
Figure~\ref{fig:figure2}(a) compares the surface-density maps at the end of the calculations ($t = 200~\mathrm{Myr}$). 
The overall gas structures (i.e., offset shocks and nuclear ring) are similar in these two models.
However, when the motion is limited to two dimensions, essentially no gas reaches the CND region ($r < 50~\mathrm{pc}$). 
To clarify the origin of the CND gas, we selected all gas particles located within $r<50~\mathrm{pc}$ at $t=200~\mathrm{Myr}$ and traced their positions back to $t=150~\mathrm{Myr}$. Figure~\ref{fig:figure2}(b) shows the radial distribution of these particles at $t=150~\mathrm{Myr}$. The blue histogram represents particles that were already inside the CND at that time, whereas the red histogram represents particles that were outside the CND at $t=150~\mathrm{Myr}$ but became part of the CND by $t=200~\mathrm{Myr}$. In the pseudo-2D model, almost all of the particles that later make up the CND are already distributed inside the ring radius at $t=150~\mathrm{Myr}$, indicating that the CND growth over this period is supplied mainly by gas residing interior to the ring. By contrast, in the isothermal-3D model, the red distribution extends well beyond the ring radius, showing that a significant fraction of the gas incorporated into the CND originates from outside the ring. This result indicates that, unlike in the pseudo-2D case, the ring in the isothermal-3D model does not completely block inward transport, and gas from larger radii can reach the central region and join the CND. 
Note that \citet{Perez2007} compared 2D and 3D gas dynamics in a bar potential and suggested that vertical effects were negligible.
However, their model probably lacked the resolution to capture the fine-scale motion near the nuclear shocks and rings. }

\begin{figure}[htb!]
\centering
\includegraphics[width=0.48\linewidth]{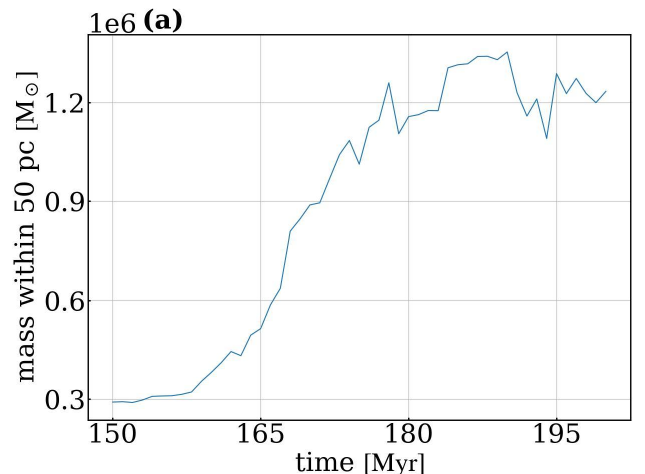}
\hfill
\includegraphics[width=0.48\linewidth]{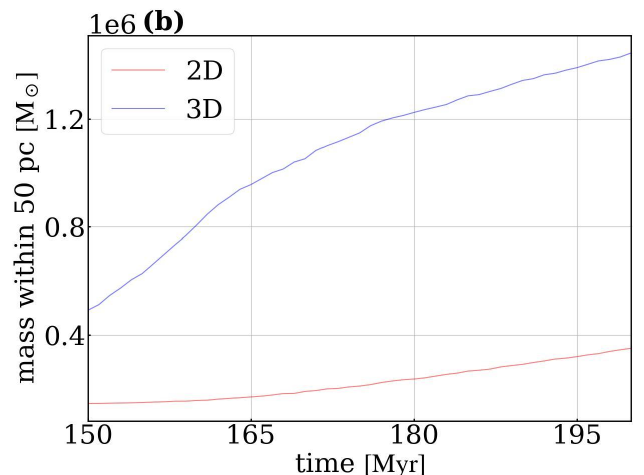}
\caption{
(a) Time evolution of the CND mass ($r < 50~\mathrm{pc}$) in the fiducial calculation that includes heating, cooling, and star formation.
(b) { CND mass evolution in the isothermal-3D and pseudo-2D models. Self-gravity of the gas is solved in both models. The blue and red curves correspond to the 3D isothermal and pseudo-2D runs, respectively.}
}
\label{fig:figure3}
\end{figure}

Figure~\ref{fig:figure3}(a) presents the CND mass as a function of time in the fiducial 3D model. About $9\times10^{5}\,M_\odot$ of gas flows into the CND between $t=150$ and $200~\mathrm{Myr}$, i.e., the average accretion rate estimated by the CND mass is $\sim 0.02 M_\odot$ yr$^{-1}$ (maximum rate is $\sim 0.17  M_\odot$ yr$^{-1}$ )\footnote{{ The average accretion rate does not show significant dependence on
on the number of SPH particles between 3 and 5 million.}}, after the bar has fully developed.
 { In the fiducial model, the increase in the CND mass appears to stall after $t \sim 180$ Myr. This is partly because angular momentum transfer by the bar becomes less effective once the elongated gas orbits align with the bar potential. The other reason is that star formation within the CND balances the mass inflow rate driven by the vaulting mechanism;
The star formation rate at $t \gtrsim 180$ Myr is $\sim 0.04 M_\odot$ yr$^{-1}$.
}

Figure~\ref{fig:figure3}(b) shows the corresponding time evolution for the isothermal 3D and pseudo-2D comparison runs. While the fiducial and isothermal 3D models continue to feed the CND, the increase of the central mass in the pseudo-2D model is much smaller than in the isothermal 3D model. 
Note that this small mass increase is not directly caused by the mass supply from the galactic disk, but by a gradual increase in the ring width.


\subsection{Physical Mechanisms Distinguishing Gas Reaching the CND from Gas Trapped in the Ring}

In this subsection we sample SPH particles that either remain in the nuclear ring or flow into the CND and analyze their statistical properties. We find that gas reaching the CND avoids the ring by {\it detouring above and below the disk plane}. These excursions arise from oblique shocks that develop in the high-density regions, including the offset shocks (observationally known as ``dust lanes'' in barred galaxies), outside the ring. The separation between the gas particles that remain in the ring and those that reach the CND already manifests outside the ring ($R > R_{\rm ring}$), indicating that access to the CND is determined by the angular momentum and scale height of the gas before it encounters the ring.

\subsubsection{Differences in the $L_z$--$z$ Plane and the Origin of Vertical Motion}
\label{sec:vertical}

To characterize the differences between gas that flows into the CND and gas that remains in the ring, we sample gas particles according to the definition of CND and ring gas described in Section \ref{sec:2.4} yields 8316 CND particles and 17,439 ring particles. To enable a direct statistical comparison, we randomly select 8316 of the ring particles so that both samples contain the same number of members.

\begin{figure}[htb!]
\centering
\includegraphics[width=170mm]{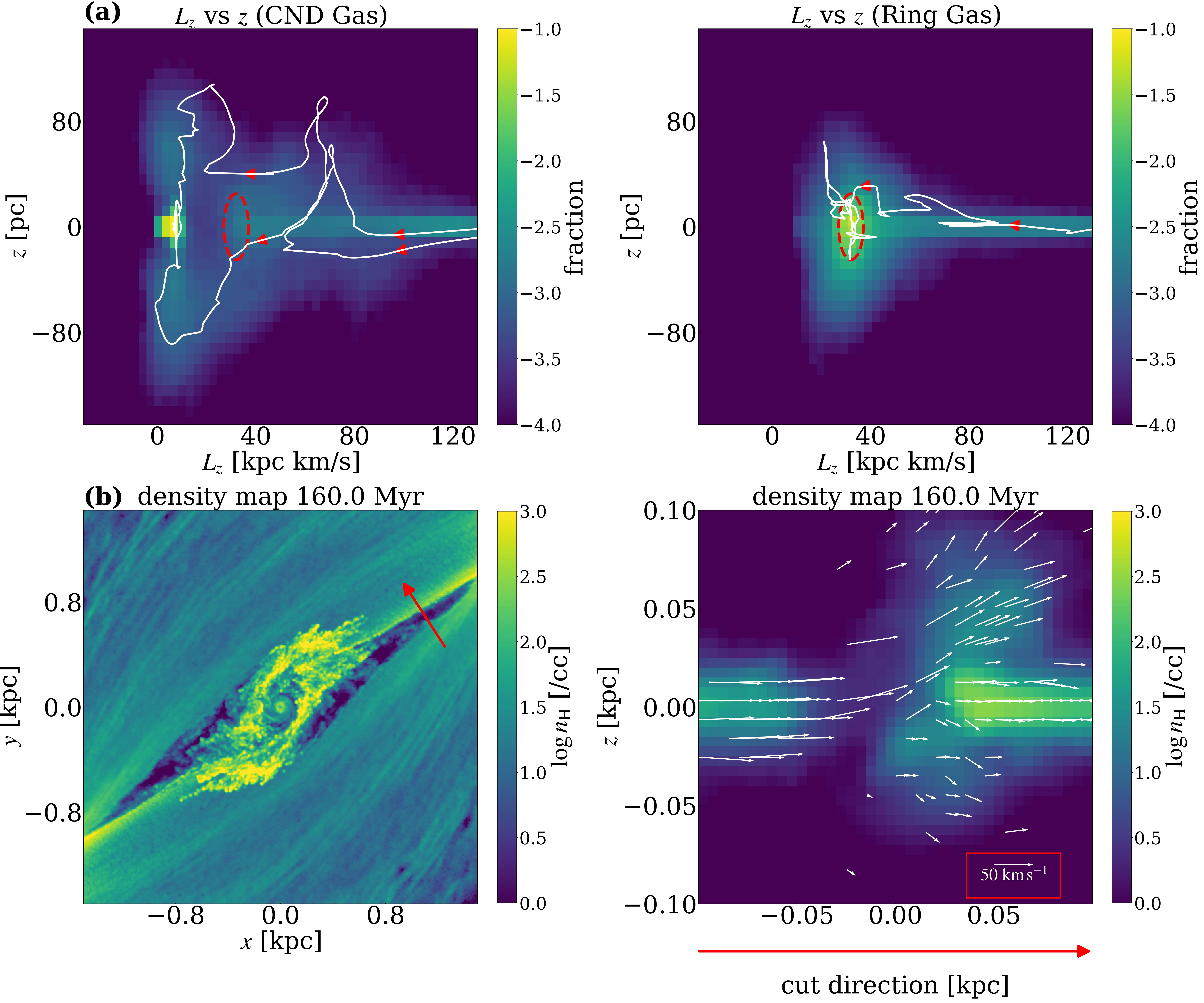}
\caption{
\textbf{(a)} Heat maps of the orbital distribution in the $L_z - z$ plane for CND gas (left) and ring gas (right). The color scale shows the particle fraction in each bin (logarithmic scale) and only bins with a fraction $> 1\times10^{-4}$ are plotted. White lines trace representative particle trajectories. Red arrows indicate the direction of particle motion. The red dashed ellipse indicates the nuclear ring region.
\textbf{(b)} Left: face-on map of the mean hydrogen number density in the dust-lane region. Right: vertical slice along the direction marked by the red arrow in the left panel; the vertical axis measures the height above the mid-plane and the white arrows indicate velocity vectors projected onto the slice. The slice averages over $\pm 100~\mathrm{pc}$ along the line of sight.
}
\label{fig:figure4}
\end{figure}

{ 
To quantify the orbital differences between gas particles that ultimately reach the CND and those that remain in the ring, Figure~\ref{fig:figure4}(a) shows
``heat maps” in the $L_z - z$ plane, where $L_z$ is the $z$-component of the specific angular momentum and $z$ is the coordinate perpendicular to the galactic plane. The color scale indicates the fraction of particles occupying each bin, and the white curves trace two representative trajectories
for the CND gas and one for the ring gas.
To highlight the main structures we only plot bins with particle fractions exceeding $1\times10^{-4}$.
}

{
Gas that reaches the CND (Figure~\ref{fig:figure4}a) occupies a broad range of angular momentum, extending from the ring value ($L_{z, ring} \sim 30~\mathrm{kpc\,km\,s^{-1}}$) to $\sim 100~\mathrm{kpc\,km\,s^{-1}}$, and reaches relatively large $|z|$. Instead of accumulating at the ring, this gas passes the ring with $L_z \ll L_{z, ring},  L_{x,y} \ne 0$ and $|z| \gtrsim 50~\mathrm{pc}$, implying that it enters the CND from polar directions.
In contrast, most ring particles remain near the mid-plane ($|z| \sim 0$) with $L_z \gtrsim 60~\mathrm{kpc\,km\,s^{-1}}$, and attain appreciable scale height only near $L_z \sim 30~\mathrm{kpc\,km\,s^{-1}}$ (the right panel of Figure~\ref{fig:figure4}a).
{When the launched gas reaches the ring region, indicated by the red dotted circle, interaction with the ring gas lifts it to larger scale heights. If it loses sufficient angular momentum, it can vault over the ring and reach the center. During this fueling process, some particles cross the disk and accrete onto the CND from both above and below the plane.}
}
The dichotomy in the fate of the gas originates in the kiloparsec-scale gas dynamics outside the ring. Figure~\ref{fig:figure4}(b) shows the density and velocity structure of the shock surfaces in the dust-lane region. Strong vertical motions accompany the shocks, indicating that they are oblique shocks. The three-dimensional geometry of these shocks launches gas above and below the galactic plane.

\subsubsection{Conditions for CND Inflow and Locations of Shock Encounters}
\label{sec:shock}
\begin{figure}[htb!]
\centering
\includegraphics[width=100mm]{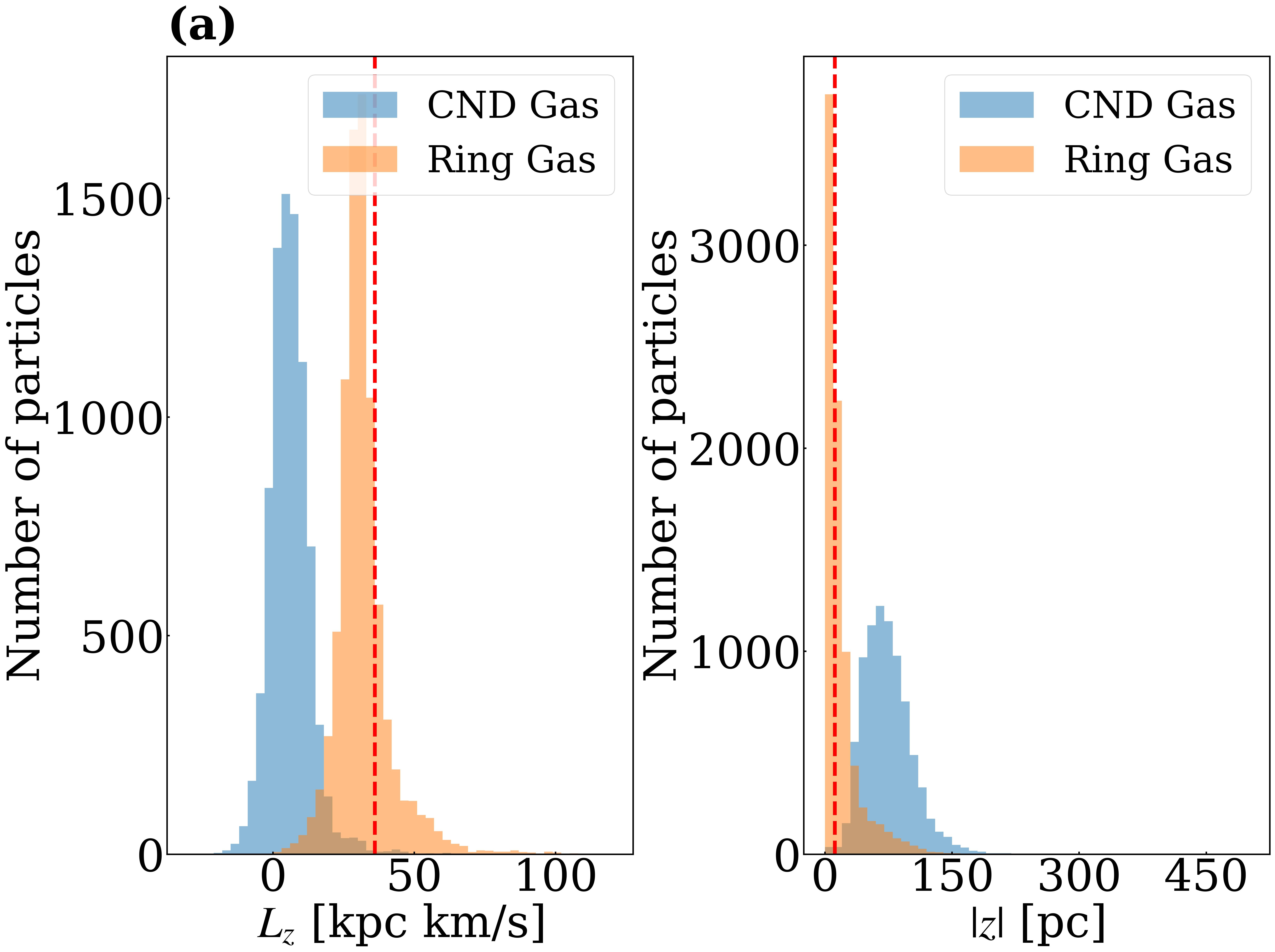}
\par\bigskip
\includegraphics[width=100mm]{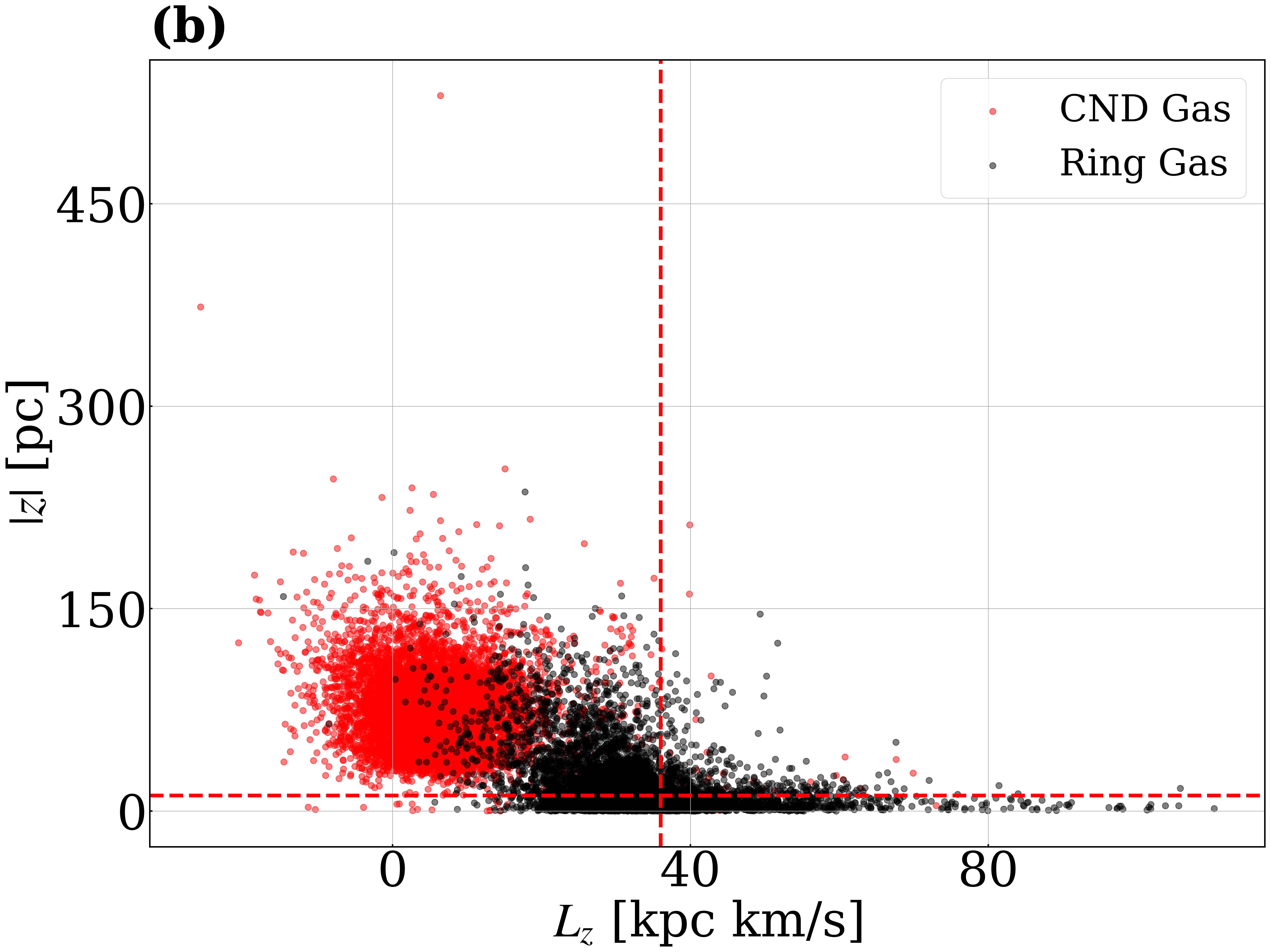}
\par\bigskip
\includegraphics[width=100mm]{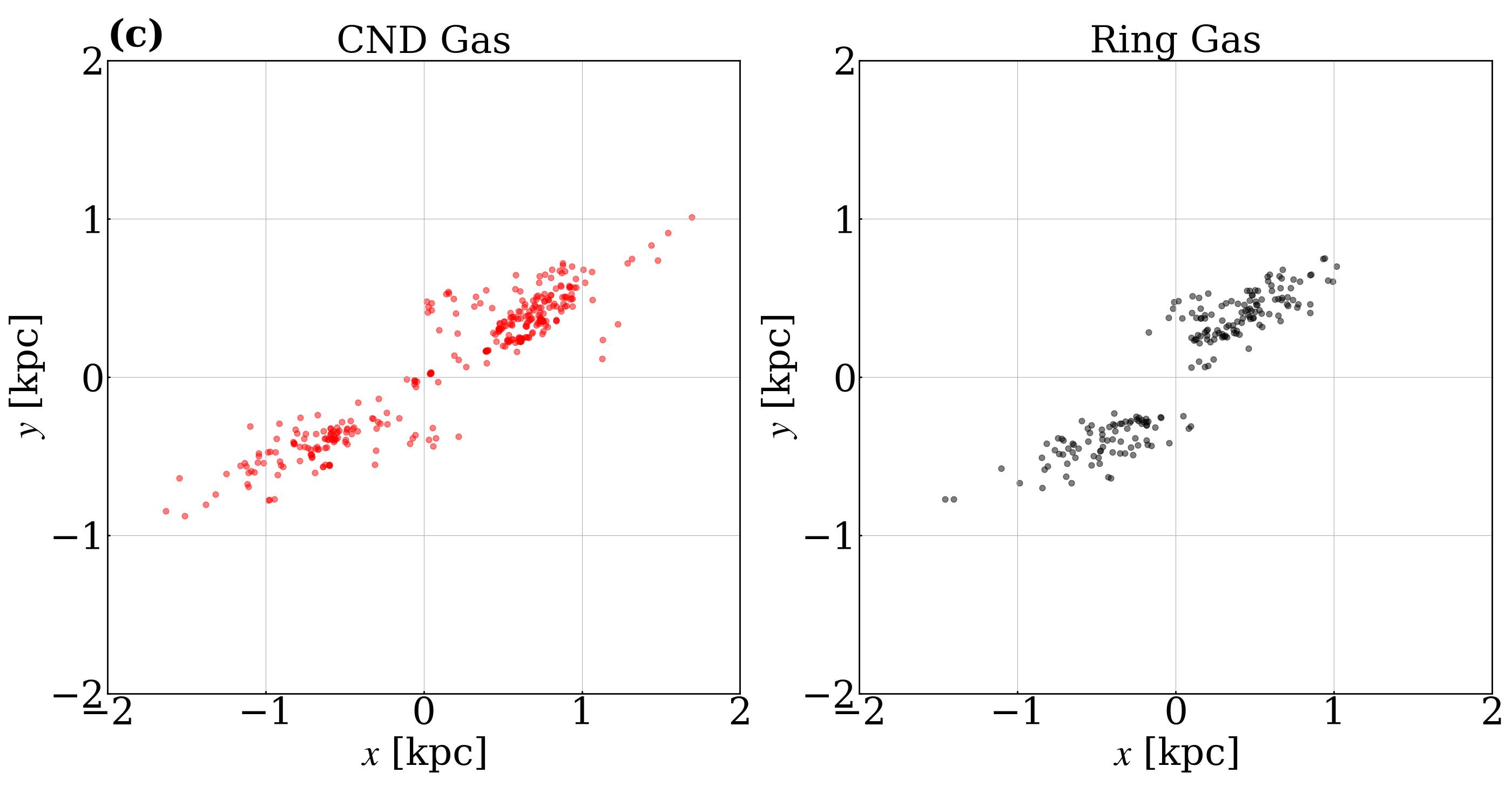}
\caption{
\textbf{(a)} Histograms of $L_z$ (left) and $|z|$ (right) when CND and ring particles last crossed $R=380~\mathrm{pc}$. The red dashed lines show the time-averaged \textit{half-mass height} and $L_z$ inside the ring region.
\textbf{(b)} Scatter plot of the same data in the $L_z$--$|z|$ plane. Red and black points correspond to CND and ring particles, respectively; the dashed lines again show the ring averages.
\textbf{(c)} Locations of events where the magnitude of the angular momentum changes by $-10~\mathrm{kpc\,km\,s^{-1}}$ or more within 1 Myr. The left and right panels refer to the CND and ring samples, respectively.
}
\label{fig:figure5}
\end{figure}

Figure~\ref{fig:figure5}(a) shows histograms of $L_z$ and $|z|$ for CND and ring particles at the moment they {\it last} crossed $R=\sqrt{x^{2}+y^{2}}=380~\mathrm{pc}$, i.e., just outside the nuclear ring\footnote{The time-averaged ring radius is $R_{\rm ring} = 277.5~\mathrm{pc}$.}. 

Figure~\ref{fig:figure5}(b) plots the same data in the $L_z$--$|z|$ plane. The dashed lines mark the time-averaged \textit{half-mass height} and $L_z$ within the ring region ($0.7\,R_{\rm ring} < R < 1.3\,R_{\rm ring}$).  These plots show that the two populations are already {\it separated outside the ring}. Ring particles retain $L_z$ and $|z|$ values comparable to those of the ring itself, whereas CND particles exhibit significantly lower $L_z$ and larger $|z|$. Only the subset that meets these conditions can avoid capture by the ring and plunge from above or below into the CND.

Figure~\ref{fig:figure5}(c) locates the ``angular-momentum loss events,'' defined as timesteps where the hydrodynamic torque (Section~\ref{sec:2.5}) changes the angular-momentum magnitude by
\[
\Delta L_{\mathrm{hydro}}^{\Delta t = 1\,\mathrm{Myr}} < -10~\mathrm{kpc\,km\,s^{-1}}
\]
within a 1 Myr interval. Both CND and ring particles predominantly lose angular momentum in the offset shocks and the ring itself, consistent with the regions where shocks form. These losses are therefore mainly shock-driven. CND and ring particles experience shocks in nearly the same locations, but the outcomes differ: shocks reduce the total angular momentum $|L|$ of both groups, yet ring particles primarily lose $L_z$, causing them to settle near $z=0$ and remain in the ring (Figure~\ref{fig:figure5}). CND particles, on the other hand, gain $L_x$ and $L_y$, tilting their angular-momentum vectors and increasing $|z|$, and continue to fall inward while $L_z$ keeps decreasing. They thereby vault over the ring and reach the center. The contrasting distributions in the $L_z$--$|z|$ plane (Figure~\ref{fig:figure5}) capture this difference. Only a minority of gas with substantial angular-momentum loss and appreciable vertical motion avoids capture by the ring and contributes to the CND.

\section{Discussion}

\subsection{Comparison with recent hydrodynamic simulations of the Galactic center.}

{   In our simulations, we have identified a novel intrinsic fueling channel, the ``vaulting mechanism", in which a fraction of the gas flowing along the offset shocks (dust lanes) acquires substantial vertical momentum, allowing it to bypass the nuclear ring and directly reach the circumnuclear disk (CND). Recently, a series of high-resolution 3D hydrodynamic and magnetohydrodynamic (MHD) simulations have extensively investigated the gas dynamics around the Central Molecular Zone (CMZ) of our Galaxy  \citep{ Sormani2018, Sormani2019,  Tress2020, Tress2024}.
While our model shares several basic physical ingredients with these studies—such as 3D gas dynamics, radiative cooling, and an external barred potential—the resulting gas flows in the central few tens of parsecs show notable differences.
Specifically, \citet{Tress2020} and \citet{Tress2024} compared their models---which incorporate stellar feedback and magnetic fields---with the corresponding models from earlier non-self-gravitating simulations without star formation \citep{Sormani2018, Sormani2019}. By comparing these models, they highlighted that in the absence of stellar feedback or magnetic fields, the bar-driven inflow entirely stalls at the nuclear ring, leaving the central 50 pc region devoid of gas.
 Furthermore, \citet{Sormani2019} specifically searched for vertical ``hydraulic jumps" near the dust lane shocks to explain the vertical thickness of observed clouds in our Galaxy, but found no evidence of such large-scale vertical motions in their simulations.
 }
 
 { Because the numerical methods and gravitational potentials are not identical, no definitive conclusions can be drawn. However, several factors may account for the differences between our results and those of \citet{Sormani2018}, \citet{Sormani2019}, \citet{Tress2020}, and \citet{Tress2024}. One important difference is the adopted scale height of the disk component. In our simulations, we use a Miyamoto--Nagai disk with a scale height of $b = 400$ pc
  (Equation~(\ref{eq: MN-disk})), whereas the corresponding value in \citet{Tress2020} and \citet{Tress2024} is smaller, about 200 pc. A smaller scale height implies a deeper gravitational potential well in the $z$-direction, making it harder for gas to acquire significant vertical motion.}
{ To quantify this effect, we carried out an additional simulation with a scale height of 200 pc. We did not find a significant difference in the mass accretion rate to the CND relative to the fiducial model. We also ran the same 3D isothermal model with a scale height of 200 pc, and found that the mass accretion rate to the central 50 pc is about 60\% of that in the model presented in Fig.~\ref{fig:figure3}.}
{ In addition, the bar strength, the resonance locations (through the pattern speed), and the initial gas density distribution may all influence the efficiency of the vaulting mechanism; a systematic parameter study is left for future work.}

{ We also note that the absence of stellar feedback and the limited resolution in our model make the dust lanes appear relatively smooth compared to those in \citet{Tress2020}. Nevertheless, as illustrated in the movie accompanying Figure \ref{fig:figure1}, the gas flow in the fiducial model is not entirely smooth, unlike that in the 2D isothermal model (Fig.~2a). Our analysis suggests that inhomogeneities in the shocks and the ring also contribute to the vaulting mechanism. Specifically, not all gas parcels lose angular momentum at the shock in the same way: some encounter a denser part of the shock, lose a large amount of angular momentum, remain at a small scale height, and consequently fall directly into the ring. Other parcels lose little angular momentum at their first shock encounter, but acquire $z$-momentum, overshoot the ring, encounter the opposite shock, and are redirected onto more radial trajectories, allowing them to enter vaulting orbits. This diversity in trajectories---most gas settling into the ring while a fraction reaches the CND via vaulting orbits---appears to be a consequence of flow inhomogeneity and nonlinear effects.}

{ \citet{Sormani2019} searched for an effect analogous to the ``hydraulic jump" presented in this paper in an attempt to explain the unusually high vertical thickness of molecular clouds in the centre of our Galaxy (e.g. "Bania's Clump 2"). These clouds are believed to be located on the bar lane shocks, exactly where the hydraulic jump should occur. However, they could not find any evidence of vertical motions that could explain the vertical thickness of the clouds.
Could the large vertical thickness observed in the CMZ be driven by the same mechanism presented in this paper?
One should note, however, that although the present paper emphasizes the importance of vertical motions, the actual vertical excursions ($\lesssim 100$ pc) are considerably smaller than the overall scale of the galactic disk (note that the $z$-axis in the bottom panel of Figure \ref{fig:figure1} is stretched by a factor of $\sim 4$).}

{ In our simulation, the half-mass scale height of ring gas passing through $r = 380$ pc is only $\sim 10$ pc (although a fraction of the gas exceeds 50 pc). It may therefore be difficult to reproduce the thick ($\sim 70$ pc) vertical structure of the CMZ through purely hydrodynamic means, without the effects of supernova feedback or magnetic fields. Since star formation in our model occurs predominantly in the ring region, the effective scale height in a more realistic setup could be larger. Examining the spatial distribution of star-forming particles, we find that star formation occurs mainly in the 
three regions:  CND, nuclear ring and the offset shock regions. The total star formation rate averaged over $t = 180$--$200$ Myr is $\sim 5$--$6\ M_{\odot}\ \mathrm{yr}^{-1}$, of which the contribution from the nuclear ring is $\sim 2$--$3\ M_{\odot}\ \mathrm{yr}^{-1}$.
}

The simulations by \citet[][hereafter CV24]{Chaves-Velasquez2024-ty} emphasize that the inflow toward CMZ in our Galactic center is intrinsically three-dimensional. They report that gas moving {\it inward} during the late stage of ring formation has a {\it large vertical extent}, whereas gas moving {\it outward} occupies a {\it thinner layer}. This behavior matches our results: inflowing gas primarily follows the offset shocks, and as shown in Figures~\ref{fig:figure4} and \ref{fig:figure5}, the three-dimensional shock geometry imparts substantial vertical motion, particularly near the ring. Outflowing gas, by contrast, retains relatively high angular momentum and lower density, representing gas that has not yet encountered the strong shocks (or has only experienced weak shocks) and therefore remains geometrically thin.

CV24 proposed two possible interpretations for the vertical broadening: (1) breathing oscillation modes in which the disk thickness varies symmetrically about the mid-plane \citep{WaltersCox2001, WidrowBonner2015}, and (2) vertical motions excited by large-scale shocks. Our results strongly favor the latter. The phenomenon closely resembles a ``hydraulic jump'' in which gas encountering the barred or spiral-arm potential is suddenly deflected upward \citep{MartosCox1998, GomezCox2004}. Shocks in the dust lanes heat and stir the gas vertically, producing the thick inflow layer that ultimately feeds the nucleus.

\subsection{Comparison with Observations and Predictions}
Recent simulations and observations suggest that gas streaming along a bar does not halt at the entrance to the CMZ but instead overshoots the central region and later returns \citep{Sormani2020}. \citet{Su2025} report strongly tilted gas structures near dust lanes and argue that collisions between overshooting streams dissipate energy and cause the gas to settle toward the center. In our simulation some gas likewise overshoots (see the upper panel of Figure \ref{fig:figure1}). Gas that does not overshoot falls along the offset shocks with only modest growth in $|z|$, joining the ring. A fraction grazes the ring, overshoots to the opposite side, strikes another shock, and gains vertical height; when it subsequently falls back it can vault over the ring and feed the central tens pc region.

High-resolution ALMA surveys such as GATOS and NUGA have revealed that molecular tori on $\sim$10 pc scales often possess orientations that are kinematically decoupled from their host galaxies \citep{GarciaBurillo2021, Combes2019}. Traditional ``bars-within-bars'' scenarios, which assume smooth planar inflow, struggle to explain the observed lack of correlation between the torus position angle and the stellar bar or disk axis in nearby Seyfert galaxies. Our vaulting mechanism provides a natural explanation: gas reaching the circumnuclear region approaches from directions closer to the rotation axis rather than along the disk plane (Figure~\ref{fig:figure1}). Consequently, the CND angular-momentum vector can deviate strongly from the host-disk normal, with the direction determined by the instantaneous mixture of $z>0$ and $z<0$ inflows. Because these inflows are not homogeneous and vary over time, a broad distribution of orientations is expected. Future ALMA observations may identify gas that is currently ``vaulting'' near the CND; some features in the position-velocity diagram interpreted as outflows might be inward-moving streams along the rotational axis.

\section{Conclusion} \label{sec:conclusion}

Our high-resolution three-dimensional SPH simulations show that gas can reach the central tens of parsecs of barred galaxies through a purely dynamical, intrinsically three-dimensional pathway. We identify this process as the ``vaulting mechanism,'' in which a fraction of the bar-driven inflow avoids long-lived stagnation at the nuclear ring and instead reaches the circumnuclear disk (CND). The gas that fuels the CND originates outside the nuclear ring, rather than being supplied by secondary accretion from it. By acquiring large vertical motion ($|z| \sim 100$~pc) and losing much of its azimuthal angular momentum ($L_z$), the gas effectively vaults over the ring and plunges into the central tens-of-parsecs region.

This fueling channel disappears in our pseudo-2D comparison runs, in which vertical motions are suppressed and the gas stalls at the nuclear ring. In this sense, even in geometrically thin disks, ordered three-dimensional motions play an essential role in delivering gas to the galactic center. The required vertical excursions are generated by oblique shocks in the dust-lane regions outside the ring, in a process reminiscent of hydraulic jumps in spiral arms \citep{MartosCox1998, GomezCox2004}. 
{ The jumps also occur when the radially inflowing gas interact with the ring gas.}
These shocks convert part of the in-plane kinetic energy into vertical motion, while the associated hydrodynamic pressure gradients redirect the angular momentum vector in three dimensions. At the same time, because the flow evolves in a barred potential, gravitational torques from the bar act together with hydrodynamic forces to reduce the gas angular momentum, especially $L_z$, thereby enabling infall across the ring.

Because the vaulting gas approaches the center from above or below the disk plane, the resulting CND can acquire an angular momentum vector significantly misaligned with that of the host galaxy. This dynamically driven three-dimensional fueling mechanism provides a natural explanation for the kinematically decoupled molecular tori revealed by recent high-resolution ALMA observations \citep[e.g.,][]{GarciaBurillo2021, Combes2019}. It also offers a physical framework for the coexistence of AGNs and nuclear starburst rings: the ring acts not as a complete barrier, but as a dynamical filter that allows only a small, yet sufficient, fraction of the inflowing gas to reach the central supermassive black hole. 

\begin{acknowledgments}
We thank the anonymous reviewer for their thoughtful comments and suggestions.
This work was supported by JSPS KAKENHI Grant Numbers 25H00671 (KW) .
\end{acknowledgments}
\bibliography{wada_reference_all_v014}
\bibliographystyle{aasjournal}
\end{document}